Non-isothermal physical and chemical processes in superfluid helium

E. B. Gordon[1*], M.I. Kulish[1], A.V. Karabulin[2], and V.I. Matyushenko[3]

[1]Institute of Problems of Chemical Physics, Russian Academy of Science, 142432, Chernogolovka, Moscow Region, Russia;

[2]National Research Nuclear University MEPhI (Moscow Engineering Physics Institute), 115409, Moscow, Russia.

[3]The Branch of Talrose Institute for Energy Problems of Chemical Physics, Russian Academy of Science, 142432, Chernogolovka, Moscow Region, Russia.

Abstract

Metal atoms and small clusters introduced into superfluid helium (He II) concentrate there in quantized vortices to form (by further coagulation) the thin nanowires. The nanowires' thickness and structure are well predicted by a double-staged mechanism. On the first stage the coagulation of cold particles in the vortex cores leads to melting of their fusion product, which acquires a spherical shape due to surface tension. Then (second stage) when these particles reach a certain size they do not possess sufficient energy to melt and eventually coalesce into the nanowires. Nevertheless the assumption of melting for such refractory metal as tungsten, especially in He II, which possesses an extremely high thermal conductivity, induces natural skepticism. That is why we decided to register directly the visible thermal emission accompanying metals coagulation in He II. The brightness temperatures of this radiation for the tungsten, molybdenum, and platinum coagulation were found to be noticeably higher than even the metals' melting temperatures. The region of He II that contained suspended metal particles expanded with the velocity of 50 m·s$^{-1}$, being close to the Landau velocity, but coagulation took place even more quickly, so that the whole process of nanowire growth is completed at distances about 1.5 mm from the place of metal injection into He II. High rate of coagulation of guest metal particles as well as huge local overheating are associated with them concentrating in quantized vortex cores. The same process should take place not only for metals but for any atoms, molecules and small clusters embedded into He II.

*Corresponding author: Gordon@ficp.ac.ru

# 1 Introduction

Until recently it was believed that superfluid helium (He II) is the simplest homogeneous low-temperature medium for guest particles suspended inside it. It means that the motion of these particles in He II should have a character of simple diffusion, and any physical and chemical processes between them should be strictly isothermical [1,2]. Indeed, such a quantum fluid as liquid helium can be regarded as a continuous medium with spatially averaged characteristics; and in its superfluid state the Helium possesses the record high, quantum thermal conductivity, which should eliminate any local overheating [3].

Thus all researchers embedded the guest particles into superfluid helium considered the processes inside He II as strictly isothermal [4-6]. This opinion have been also shared by all the authors, introducing impurity particles into small droplets of superfluid helium [2,7,8].

However, in reality He II is proved to be extremely complex and specific template for the physical and chemical processes of guest particles. First, it was found [9] that these particles tend to concentrate in the cores of quantized vortices always existing in He II. These vortices are practically one-dimensional objects with a diameter of about 1 Å and length, which can reach values of many centimeters [9]. It has recently become clear that due to the colinearity of velocities for particles captured in the vortex cores the probability of their collisions there and of subsequent chemical and physical processes is much higher in the vortices than in the bulk [10]. Moreover, the elongation of particles during their coagulation leads to increase of time that the resulting cluster spends in the vortex, and hence, to increase of their local concentration there [10]. This previously unknown rapid process of spatially inhomogeneous condensation should produce in this way the long thin filaments [11]. The results of experiments with different metals and alloys introduced to bulk He II support this conclusion [12,13]. (It was recently shown that the same behavior demonstrated the metals captured inside enough large superfluid helium droplet) [14,15]. However, it was expected that due to high thermal conductivity of He II the atoms or small clusters should adhere to each other tightly, forming either monoatomic chains or nanowires with loose fractal structure. Anyway, the nanowires grown in our experiments fortunately were dense, with almost crystalline packing and "large" diameters of a few nanometers, being close to optimal for many chemical and physical applications [16-18]. Such a behavior was associated with the existence of the limiting heat flux above which the strong turbulence develops in He II. This turbulence disrupts the laminar motion of the normal component, which is responsible for high rate of heat transfer. For the objects larger than one micron, this effect is known and the threshold heat flow is about 3 watts per cm$^2$ [19]. For very small objects the concentration of vortices is too small to disturb the flow laminarity [20]. In this case, the maximal heat transfer rate will be determined as the product of the normal component density, the temperature and the speed of laminar flow restricted by the velocity of second sound in He II, i.e.

$$W = n_n v_s k_B T, \qquad (1)$$

where $n_n(T)$ – is the density of the normal component, $v_s$ – is the velocity of the second sound, equal to $2\times10^3$ cm/s, and $k_B$ is Boltzmann constant.

The temperatures of typical experiments on impurity particles introduction in He II are T = 1.6-2.0 K, thus the density of the normal component is about 20-50% of liquid helium density, nn = $(0.4 – 1.0)\times10^{22}$ cm$^{-3}$. Therefore, the limiting heat flow should be as large as $10^3$ W/cm$^2$ for particles that small. For particles of mesoscopic size the limit of the heat flow is still unknown. But it should be compared with the heat removal rate required, for example, to prevent melting of the merging product of two metal balls with 1 nm-diameter each (as it shown [16] the clusters of such size are the "bricks" for nanowire growth in the vortex core). This rate estimates as $10^5$ W/cm$^2$ [16]. We believe that even for nanometer metal clusters the limiting heat flow will be as well significantly lower than that enormously high estimate. (At the same time in clusters consisted of few atoms the total energy released during coagulation is not sufficient to form a helium gas cavity insulating heat transfer to the liquid; this item requires special consideration which goes beyond the scope of this paper).

Under above assumptions, the scenario of events taking place in excess of the limiting heat flow is more or less clear: just after the act of coagulation the He II envelope of merging particles converts to the normal He I, it then evaporates to form the sheath filled with low-density helium gas, which reliably insulates the hot core.

The experimental results support our claim, that the coagulation of metal nanoclusters in He II is accompanied by their fusion. Indeed the dense packing of atoms in the nanowires "large" (in comparison with vortex core's thickness) diameter, and the presence of metallic spheres with the perfect shape and atomically smooth surface in the products were observed [16]. Our model of nanowire formation in He II based on the assumption of adiabatic regime of the coagulation process that leads to the cluster melting [16] has been recognized in several independent studies [21,22]. Nevertheless, we decided primarily to perform the systematic comparison of diameters of the nanowires grown in a low temperature experiments from various metals with those predicted for them by the model [16].

However, by itself, the claim that, for example, tungsten, possessing the highest melting temperature among all metals, may melt inside the liquid, which cooled down practically to zero absolute temperature and also possesses a record-high thermal conductivity, cannot but cause a certain share of skepticism from any scientist. Therefore, we have set as a major goal of this work to obtain direct experimental evidence of heating metal clusters during their coagulation in He II to high, equal to several thousand degrees temperature.

The logic of our experimental study was as follows. According to our model the nanowires are formed by fusing the spherical clusters with diameter not less than 1 nm. In such clusters the metallic binding must already exist [23] and taking into account high temperatures of clusters the density of free electrons in them must be very high.

This means that the intensity of thermal electromagnetic emission should be close for such clusters to that defined by Planck formula. Liquid helium is optically transparent, and the temperature of cryostat walls is very low, so that the corresponding thermal radiation can in principle be detected experimentally.

It should be emphasized that the process of nanowires growth in quantized vortices of superfluid helium consists of several non-isothermal stages and it cannot be characterized by a single temperature. Indeed, according to our mechanism (see [16]), the first stage is the formation of small clusters. At this stage the condensation process is not obviously adiabatical and, moreover, the clusters are still not metals and hence there are no free electrons in them. Thermal radiation should be very weak and the first stage should manifest itself as a delay of glowing. The second stage starts when the clusters acquire metallic structure; the clusters are very hot and intensively emit the light. The further coagulation diminishes the temperature of molten cluster and eventually its temperature drops down to melting temperature. Such clusters could not melt anymore and they may only fused to each other resulting in appearance of nanowire fragments directed along the vortex core, that fragments then will attach to each other and so on; this stage may be characterized as the third one. Thus, the intensive thermal emission is possible only in the second, "bright" stage, and during this stage the clusters of different size possess different temperatures. Thus, one cannot speak about real temperature of this stage, but only judge how close is the effective temperature to the melting point of the metal. The main goal of our study is not accurate measurement of temperature, but response to the question, how strong is local overheating, is it few Kelvin, or few thousand Kelvin? Accordingly, the characteristic time of coagulation process that we will determine is only the duration of the bright stage.

There is no delayed fluorescence in a metal, and the plasmon excitation by a laser lasts during the time much shorter than a nanosecond [24]. The only source of parasitic light in the microsecond time range is the radiation of plasma excited in the focal spot of the laser. Its contribution must be determined in special experiments.

**2 Experimental**

Our method for nanowires production in superfluid helium has been described in detail elsewhere [25]. The experimental setup was assembled on the base of optical liquid helium cryostat; lowering of the temperature was carried out by pumping out liquid helium vapor down to pressure of 700 Pa which corresponded to a temperature of 1.55 K. The atoms and small clusters of metal were introduced into superfluid helium bulk by laser ablation from the surface of submerged in He II targets made of corresponding metals. The pulse-repetition solid state diode-pumped Nd: LSB laser used for the ablation has the following characteristics: wavelength $\lambda = 1,064$ μm, pulse energy $E = 0,1$ mJ, pulse duration $\tau = 0.4$ ns and repetition rate $f = 0\ 4000$ Hz. Irradiation was carried out through the sapphire windows of the cryostat, the laser beam was focused on the target surface to the spot of 50-100 microns in size. The metal particles were captured to the core of quantized vortices nucleated in the laser focus and then condensate there forming thin nanowires. These nanowires joined together forming a 3D nanoweb and fallen down to the surface of usual TEM grids placed at the bottom of cell. After warming to room temperature the grids were examined with an electron microscope JEM-2100 (JEOL company).

The design of apparatus for optical measurements is clear from Fig. 1. Hamamatsu photosensor H11526-110-NN module based on photomultiplier (PMT) equipped with gate function has been used for the emission registration [26]. The sensitive diameter of photocathode is 8 mm and its spectral sensitivity is shown in [26]. Oscilloscope Tektronix TDS 7054 and the custom made gate signal generator also have been used.

In integral experiments the photocathode of photosensor module was placed at 95 mm distance from the laser focus spot on the surface of metallic target. In spatially resolved experiments the photo camera objective with focal length 75 mm was arranged so that it provides a 2x magnification image of the region in the vicinity of laser spot. The PMT equipped with the 1 mm width slit can be moved in the plane of this image along line parallel to laser beam axis. Thus the 0.5 mm space resolution was achieved in the time-of-flight measurements.

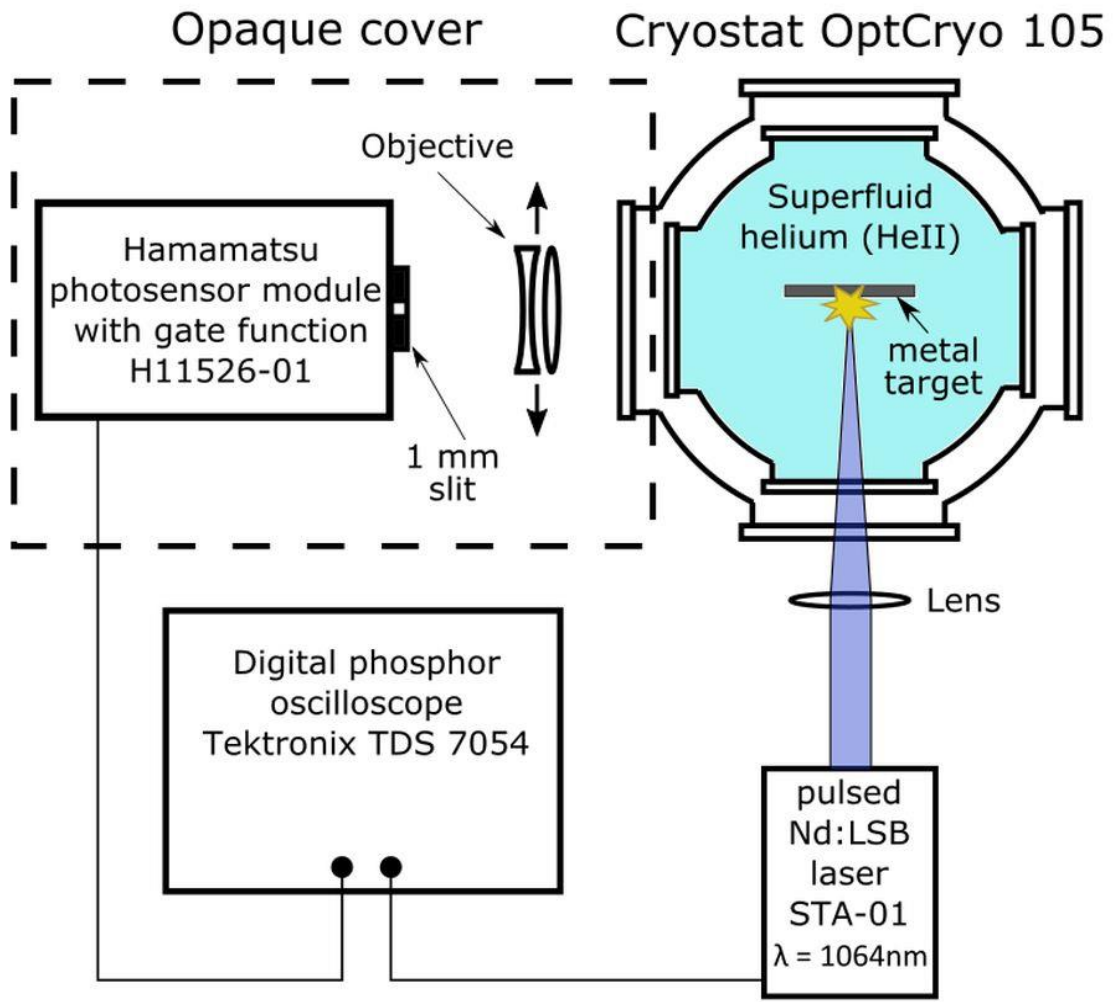
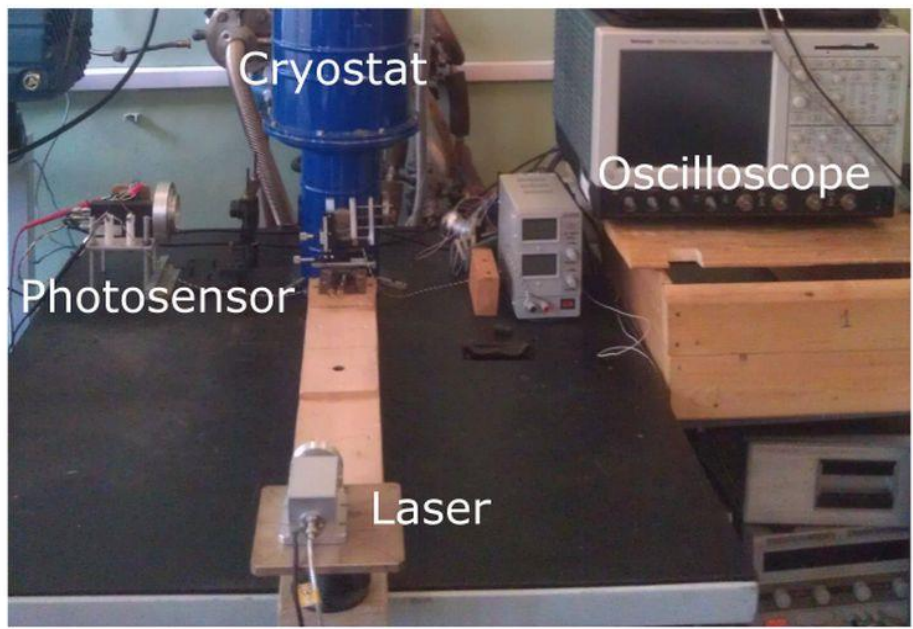

*Figure 1. The scheme and photo of the experimental setup for optical measurements.*

The gate of PMT was set by the gate generator at "OFF" ~50 μs before laser pulse. The signal to set the gate "ON" come simultaneously with laser pulse from the laser "TRIGGER OUT" and due to the internal delays both in gate generator and photosensor module the PMT was turned on at 180 ns later the laser pulse. This way we managed to avoid the saturation of the PMT by the large laser signal.

For improving signal-to-noise ratio each oscilloscope trace was averaged 128 times. The repetition rates of laser and oscilloscope trigger pulses were of 50 Hz. Laser pulses where started at zero time of oscilloscope record.

**3 The comparison of diameters of nanowires grown from different metals with the predictions of the model17**

The morphology of the samples from different metals was quite similar. They represented nanowebs with cell sides of 50-300 nm. The diameter of individual nanowires was practically the same along the whole web, but its value clearly depended on the kind of the metal. It changes from 7-8 nm for fusible indium to 2 nm for refractory tungsten. In [17] the simple formula based on thermodynamic and geometric considerations was derived

$$R_{max} = \frac{Q_b}{(C_p T + Q_m)} \cdot a \qquad (2)$$

where $Q_b$ the evaporation heat of the metal; a – the thickness of the monolayer (a ≈ 0.4 nm); $Q_m$ is the heat of melting and $C_p$ is the heat capacity of solid metal.

In this paper we compared the diameters of all nanowires grown in our studies [16,27-30] with those predicted for corresponding metals by formula (2). As it is seen from Figure 2, despite of the simplicity of the approach, there is almost quantitative agreement with all experimental data, even in absolute values of nanowire diameters.

The nanowires made of different metals had, generally speaking, different structure, some of them were single crystals, others were polycrystals, and others had amorphous structure. However, before being analyzed in the electron microscope the nanowires were heated up to room temperature and experienced contact with air. Therefore, it was unknown what structures they had in superfluid helium. Thin nanowires have proved to possess the low temperature stability [28,29], and in particular the silver nanowires which structure was carefully studied in [21] when heated to room temperature even lost their topology broking up into separeted nanoparticles. For us it is important that all of the nanowires shown in Figure 2, have dense-packing structure, which together with their diameters obeying to formula (2) are the evidences in favor of their growth via molten protoclusters.

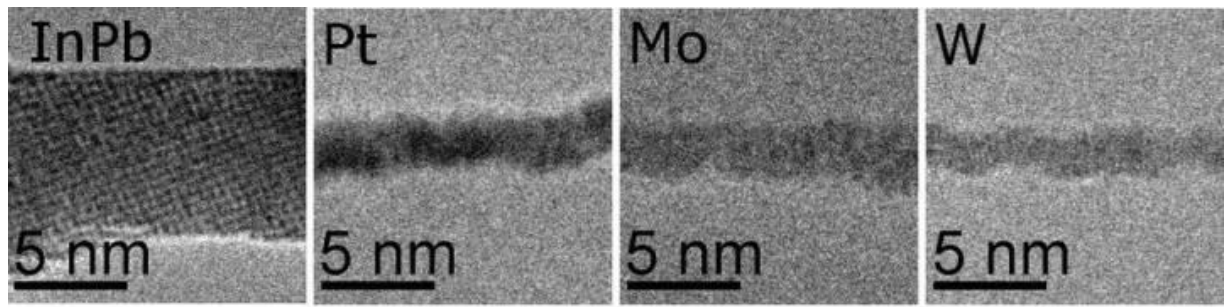

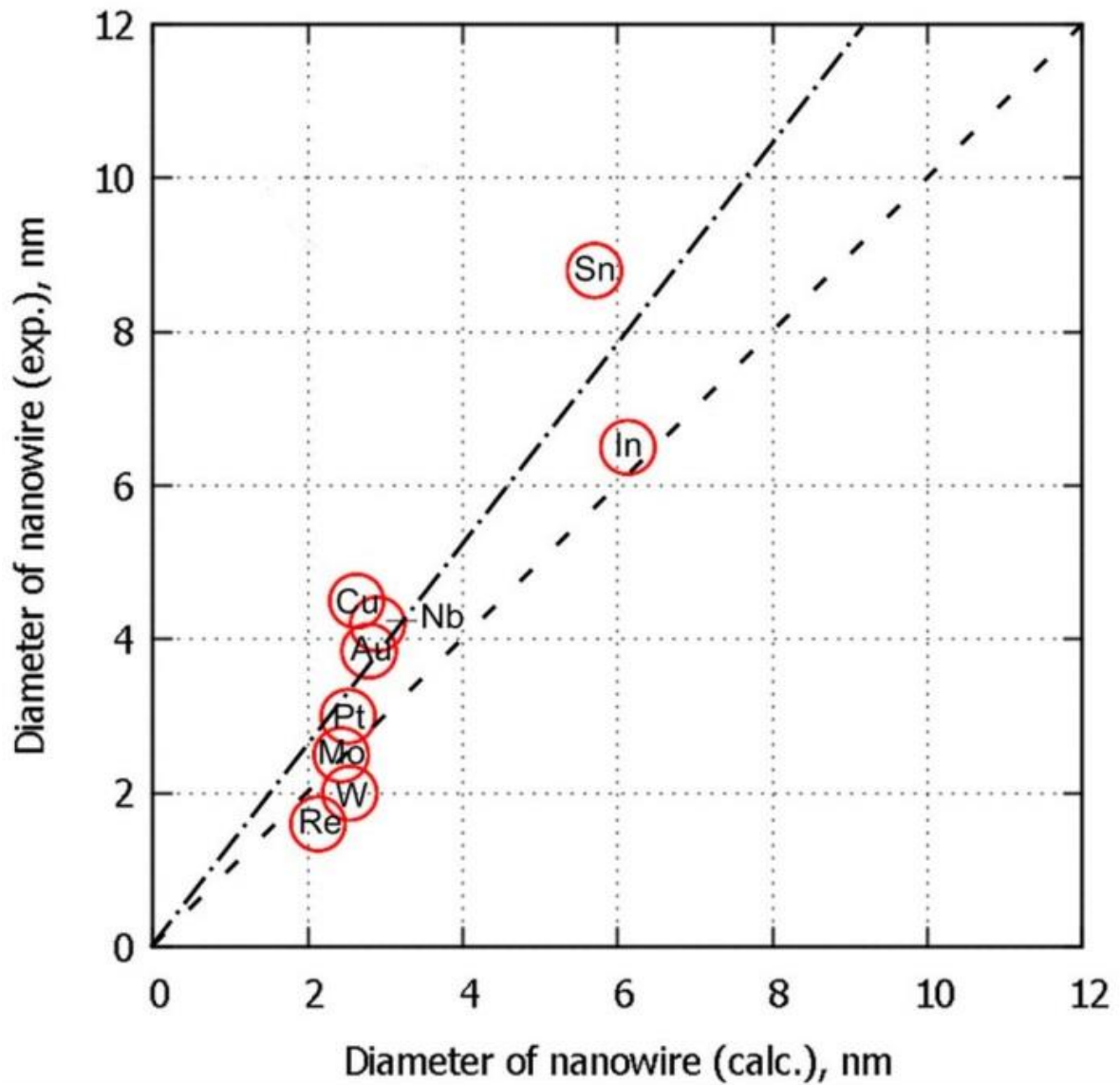

*Figure 2. The comparison of the experimental nanowires diameters with those predicted by formula (2). The structures of nanowires made of W, Mo, Pt and InPb alloy were taken from our papers [30-32].*

**4 The choice of the objects**

Our goal was to register in an optical range the thermal radiation accompanying the coagulation of metallic particles introduced into the superfluid helium by laser ablation from targets of the corresponding metal immersed in He II. The goal was to prove that the metal clusters produced by condensation were heated up to temperatures of several thousand degrees. That is why we used the photomultiplier sensitive only in visible spectrum. The choice of such metals as indium, molybdenum, platinum and tungsten as the objects of study was dictated by the goal of study; their thermal properties are listed in Table 1.

*Table 1. Some characteristics of the metals studied in this paper. $T_{melt}$ and $C_p$ are tabulated values for bulk materials [32]. $T_{ad}$ - adiabatic temperature of the 1 nm - diameter cluster formed by fusion of two cold smaller clusters, $D_{clust}$ - the diameter of the cluster formed by merging of two cold identical spherical clusters, which has a temperature $T = T_{melt}$, $\lambda_{max}(T_{melt})$ and $\lambda_{max}(T_{ad})$ - wavelengths, corresponding to maximal black-body emission for $T = T_{melt}$ and $T = T_{ad}$, correspondingly. Values marked with * obtained by using the formula 5 of paper [16].*

| Metal (atomic number) | $T_{melt}$, K | $\lambda_{max}(T_{melt})$, μm | $T_{ad}$,* K | $\lambda_{max}(T_{ad})$, μm | Heat capacity, $C_p$, J/mol·K | Latent heat of fusion ($Q_0$), kJ/mol | Full heat of melting ($Q=Q_0+C_pT_{melt}$), kJ/mol | $D_{clust}$*, nm |
|---|---|---|---|---|---|---|---|---|
| In (49) | 430 | 6.44 | 1900 | 1.53 | 26.7 | 3.24 | 11.5 | 4.78 |
| Pt (78) | 2041 | 1.42 | 3280 | 0.88 | 25.9 | 27.8 | 52.8 | 1.97 |
| Mo (42) | 2890 | 1.0 | 4200 | 0.69 | 28 | 23.93 | 80.9 | 1.89 |
| W (74) | 3695 | 0.78 | 6690 | 0.43 | 24.3 | 35.2 | 89.8 | 1.98 |

We were guided by the following considerations. According to the scenario we considered most probable [16] the nanowire starts to grow when the heat released under two metal clusters fusion becomes not sufficient to melt the cluster produced by coagulation. Therefore, at this stage the temperature of clusters should be equal to their melting temperature which is below the melting point of the bulk metal ($T_{melt}$). However, for the cluster size we are interested in, the difference is only 10-15% [33] and for generality we will neglect the difference between these two temperatures. As shown in Table 1, the maximum of black-body radiation at $T_{melt}$ for refractory Tungsten is close to the red boundary of the PMT sensitivity, so in this case the emission is easy to be registered. In contrast, the maximum black-body radiation at $T_{melt}$ for fusible Indium is far beyond the sensitivity range of the PMT, and there is no hope to register an emission during the indium coagulation.

Even for Molybdenum and especially Platinum which have intermediate melting temperatures the thermal radiation should be attenuated under registration by orders of magnitude.

However, in the early stages of coagulation the temperature of molten clusters should be much higher than $T_{melt}$. The temperatures calculated for the adiabatic merging of two spherical clusters with a diameter of 1 nm (which for all metals is less than the diameter of protoclusters) are shown in the Table 1. For such clusters the temperatures are really high, so that in the cases of Mo and Pt one can hope to register the thermal radiation corresponding to the early stages of coalescence. In the case of Indium in order to have a temperature of 3000 K the clusters would consist of only a few atoms, and that is too small to display the metallic binding, thus their emissivity should be very low.

Under this logic if the effective temperature of metallic clusters will be close to the melting temperatures of metals they made of, the large signal of thermal emission will be detected only in the case of Tungsten. But if the effective temperature of nanoclusters would turn out to be significantly higher than $T_{melt}$ the emission for Molibdenum and even for Platinum maybe detectable as well.

## 5 Optical experiments

Accordingly to existing viewpoint for ablation in liquids [35], the filled with plasma gas bubble in laser focus on the surface of target has the characteristic size of a few tenths of mm; the plasma decay products (primarily atoms and small metal clusters) embed into liquid through the bubble surface. The metal nanoparticles in He II have time to cool down before their mutual collisions; otherwise the main products of coagulation would be micron-size spheres with atomically smooth surface [16]. Ablation efficiencies for all metals under study, that measured via volume of a crater formed in the target by the laser are comparable and can be estimated as $10^{10}$-$10^{11}$ atoms per pulse with energy 0.1 mJ.

To distinguish the thermal emission of hot metal nanoclusters in the bulk of He II from the signal registered by PMT it was necessary first of all to get rid of the scattered laser radiation, that could saturate the photomultiplier dynodes by electrons for a long time. The laser pulse duration was 400 ps and the geometric dimensions of the cell could not provide the scattered light delay of more than 1000 ps. It means that the delay of PMT high voltage switching-on of 180 ns allowed to reliably avoiding this effect.

In order to remove the contribution of parasite light from plasma in the laser focus we examine separately the intensity and duration of the gas plasma glow. To do this the cryostat at the room temperature was filled with helium gas of atmospheric pressure, and the emission of the focal spot followed laser pulses was registered at 45° angle to the plane of the target.

As Figure 3 shows, the most intensive are plasma glows for indium and tungsten, in case of molybdenum the intensity was three times lower. This is quite natural, since the intensity of the metal plasma emission is mainly dependent on the density of the energy levels, the probabilities of optical transitions between them, the electron temperature, etc., but in no way it may depend on the metal melting temperature. Besides in all cases the duration of plasma emission does not exceed 1 μs.

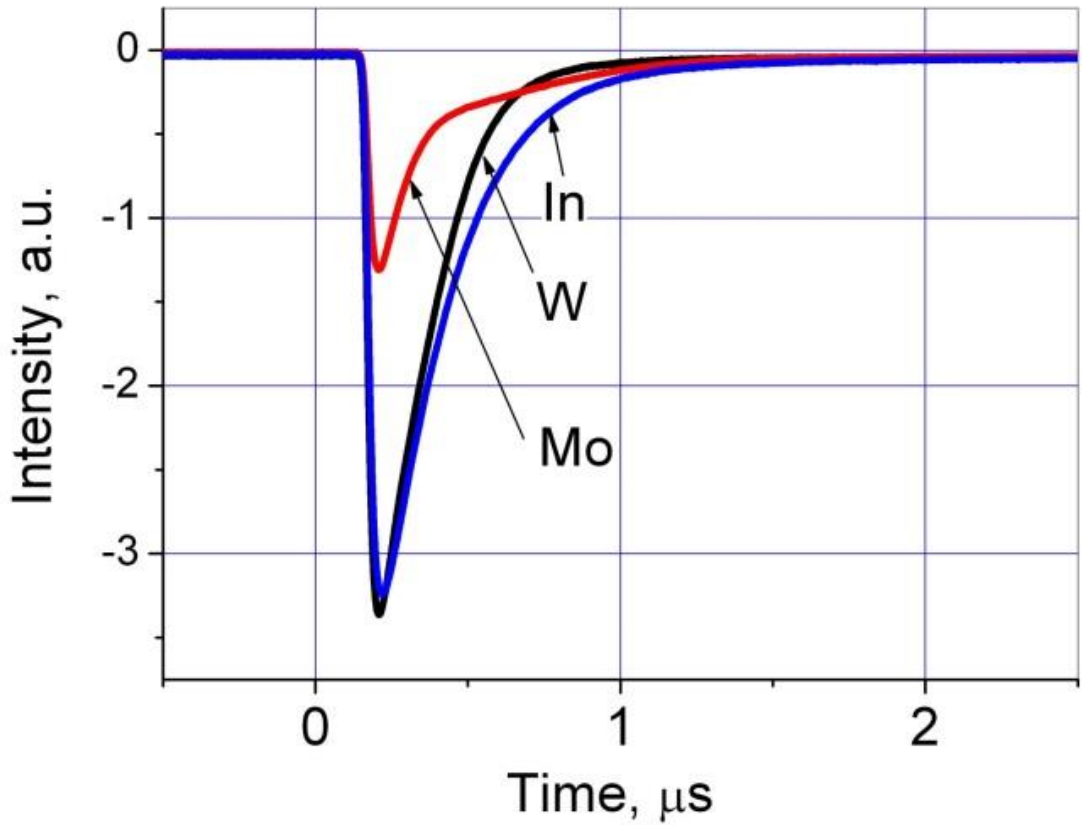

*Figure 3. Signals of emission from focal spot on the targets of indium, tungsten and molybdenum, respectively. The measurements were made at 45° angle to the plane of the target in a helium gas at room temperature and atmospheric pressure.*

The glow, accompanying the condensation of metal in superfluid helium, was registered in a plane parallel to the target plane. Therefore, the contribution of the emission of the gas bubble near the focal spot was significantly suppressed. Furthermore, as follows from Figure 3, the glow of the plasma gas could not be presented in the measured signal at times greater than 0.5 μs. Dependence of light intensity in the liquid superfluid helium induced by laser ablation of metal targets is shown in Figure 4.

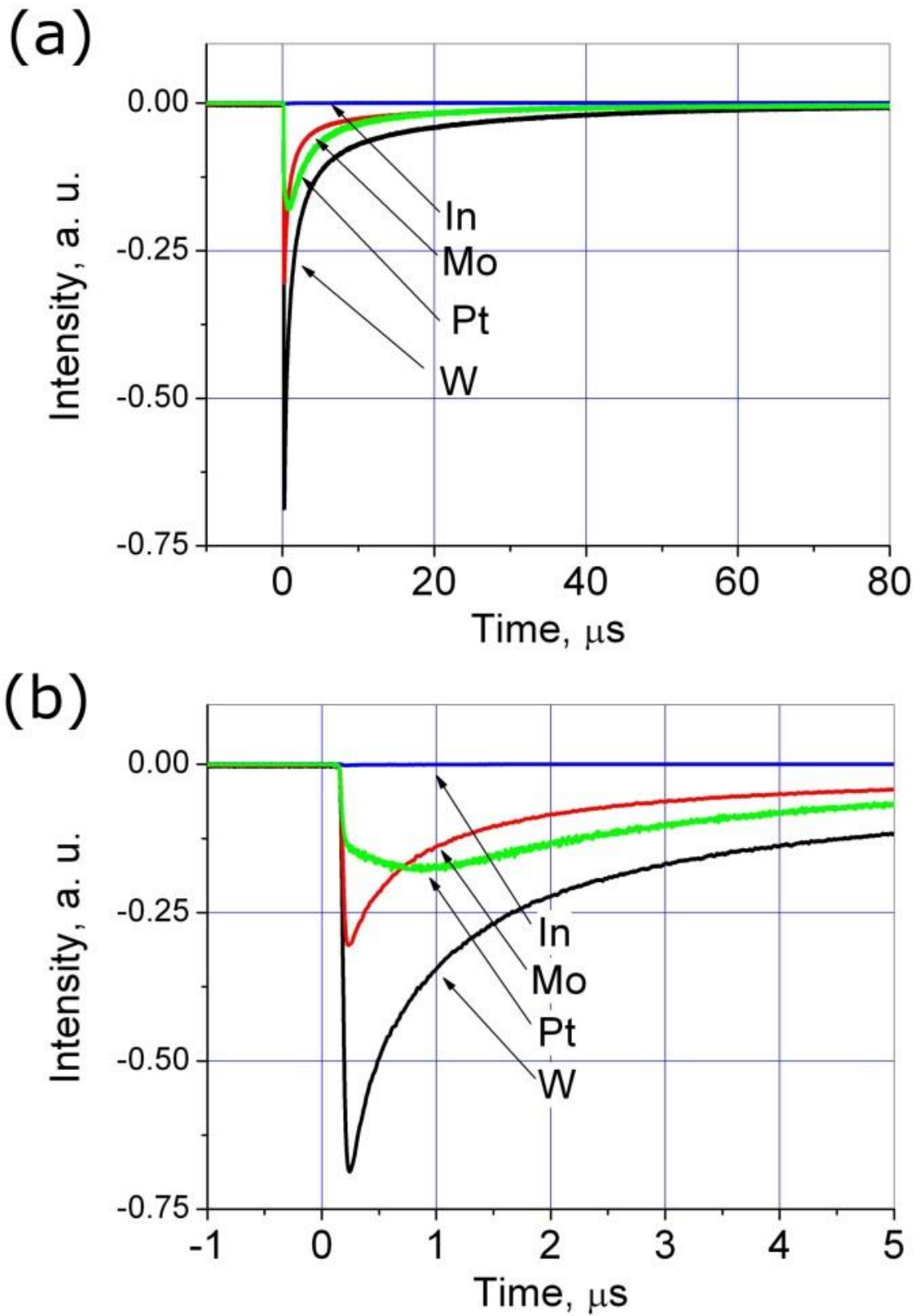

*Figure 4. The temporal behavior of visible emission in bulk He II induced by the laser ablation of indium, molybdenum, platinum and tungsten targets. Fig. 4(b) represents the front of the signals in more details.*

It is immediately seen that in case of Indium, where the laser plasma radiation has the same intensity as tungsten plasma, the emission from the bulk liquid helium is practically absent.

This is consistent with our expectations; because the estimate made with taking into account the spectral PMT sensitivity shows that the intensity of radiation accompanied the indium coagulation in superfluid helium should be at least four orders of magnitude lower than that in the case of tungsten. At the same time it served as proof of the weak contribution of plasma emission from the focal spot to observed signals.

As expected, the emission induced by tungsten coagulation is the most intensive. However, the emission for molybdenum and platinum was only a few times less intensive than that for refractory tungsten. It is the evidence that the effective temperatures for them are even higher than the melting temperature of corresponding metal.

Very important and hardly predictable feature of the emission from the bulk He II is its temporal behavior. It was suggested in [6] that the coagulation of impurity particles in the cores of quantized vortices may proceed much faster than that in the bulk superfluid helium, but it was unclear until now, what are the characteristic times of metal coagulation in quantized vortices for typical conditions of experiment. On the basis of the observed experimentally independence of grown nanowires morphology on the repetition rate of laser pulses up to 4 kHz, it has been suggested that the whole process of nanowires formation in our experiments takes less than 250 microseconds [31]. The results of present study allow giving more definite answer for the question.

As it can be seen in Figure 4, the duration of the "bright" stage of the process, though it depends on what metal is investigated, nevertheless, was always much shorter than above estimate and never exceeded 20 microseconds. Of course, according to our scenario [8] the "bright" stage of molten spherical clusters growth should be followed by the "dark" stage of their sticking together into nanowires without melting. However, due to the acceleration of the process of metal condensation in quantized vortices with the increase of condensate size, the "dark" stage of the process which consists in mutual coalescence of spherical protoclusters should proceed faster than the "bright" stage of molten spherical clusters growth. On the other hand, as it will be shown below the whole process of condensation proceeds at the distance less than 1.5 mm from laser spot, so the geometric factor in collection of the light by the PMT should have no impact on the time profile of the observed signal. However there are other factors that may change the signal profile, in particular, the process of metal nanoparticles expansion inside He II, as well as the process of hot clusters cooling which may change the spectral composition of the emission.

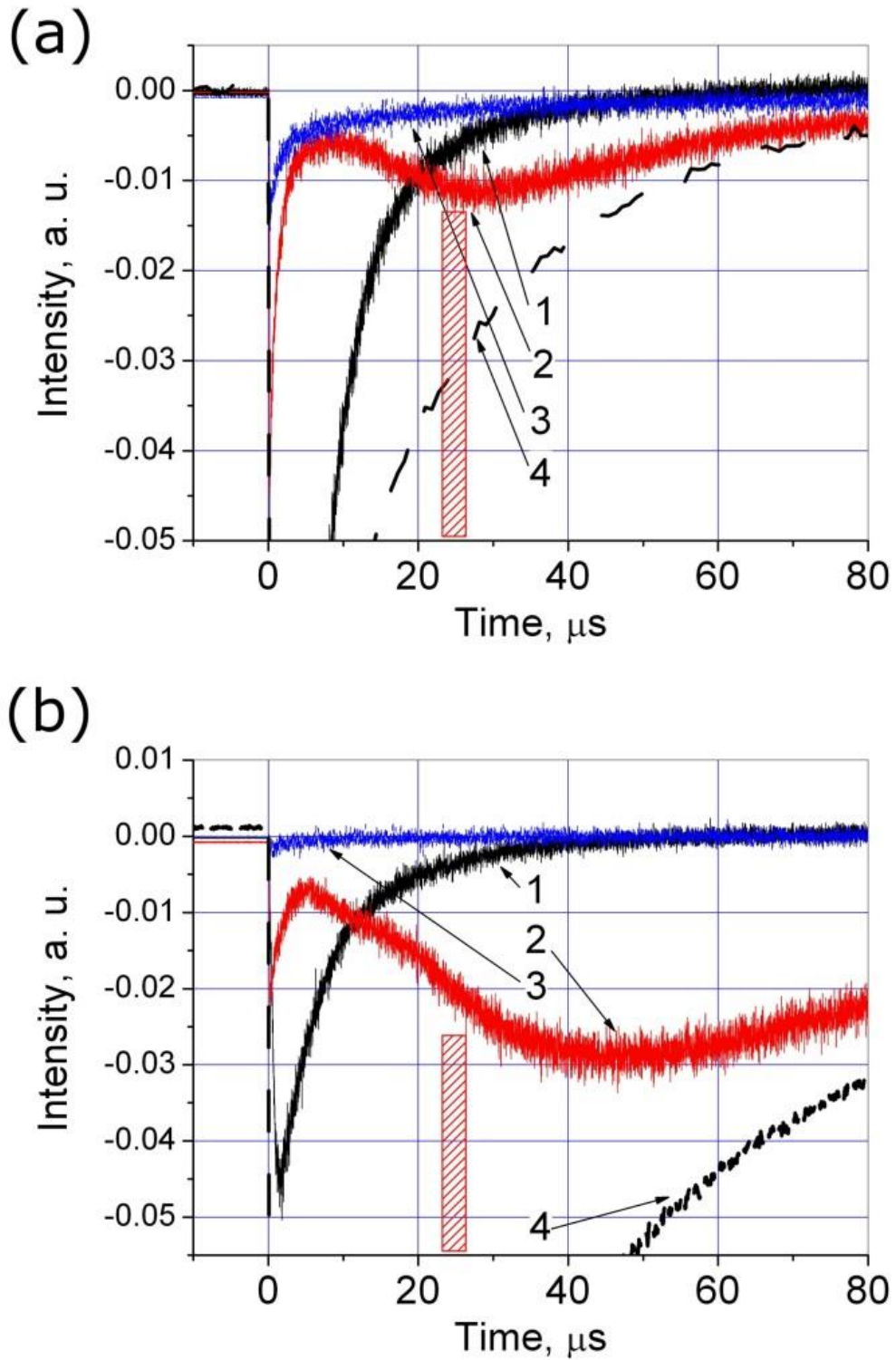

*Figure 5. The spatial distribution of the thermal emission in superfluid helium. The curves corresponded to three distances from laser focus (1) - 0, (2) - 1.25, and (3) - 2.5 mm. The rectangles mark the moment of arrival at the observation point 1.25 mm of flat front moving with Landau velocity, $V_L$ = 50 m/s. Curves 4 are the integral emission signals registered for the similar conditions in absence of optical slit). a - tungsten, b - molybdenum.*

In order to reveal the contributions of these factors to the emission signal the special experiments were carried out. Influence of metal expansion in superfluid helium on the shape of the radiation signal has been studied with the aid of spatially resolved techniques. By moving the photomultiplier equipped with a vertical optical slit with 1 mm width, along the doubly magnified image of coagulation area created by the objective 3 (see Fig. 1). The signals were recorded at the following positions of the optical slit: (i) in the place of the laser spot image, (ii) at distance of 1.25 and (iii) at distance of 2.5 mm from it.

It is clear from Figure 5 that the expansion of the region of superfluid helium filled with suspended metal proceeds very fast. Its velocities for both W and Mo are close to Landau velocity (50 m/s) which is the maximal velocity for motion of any guest particles in He II without friction. It is worth to note that rectangles in Figure 5 represent the times corresponding to the motion with Landau velocity in the direction vertical to the target – for the propagation at some angle to the vertical the time of metal appearance at a given distance from the target plane should be longer.

However, the comparison of the delayed signal 2 with integral signal 4 shows that at the distance of 1.25 mm from the focal spot the emission is already quite weak and at the distance of 2.5 mm it completely disappears (see signal 3). This means that the "hot" stage of coagulation proceeds even faster than the metal expansion inside He II and whole process of metal condensation in superfluid helium occurs in vicinity of the place of ablation.

Methodologically it is a very important result. It shows that the intensity of the laser ablation, which we used for nanowires synthesis, leads to initial concentration of the metal embedded into the He II so high that the whole coagulation process in the vortices occurs very close to the surface of gas bubble existed in laser focus. This, of course, does not promote the formation of high-quality and long nanowires.

The spectra of emission have not yet been studied in our experiments and in order to reveal how strong are the temporal changes of the radiation spectra we used the broadband glass filters, "blue" SZS20 and "red" KS10 (their transmission spectra together with the spectral sensitivity of PMT photocathode shown in Fig.6). With taking into account the spectral efficiency of the cathode one can consider that the first filter extracts 380-450 nm spectral range while the second one registers the radiation in 600-650 nm region. Fig. 6.a shows the temporal dependences of the intensity of the emission induced by the tungsten coalescence registered with the usage of "blue" (1) and "red" (2) filters.

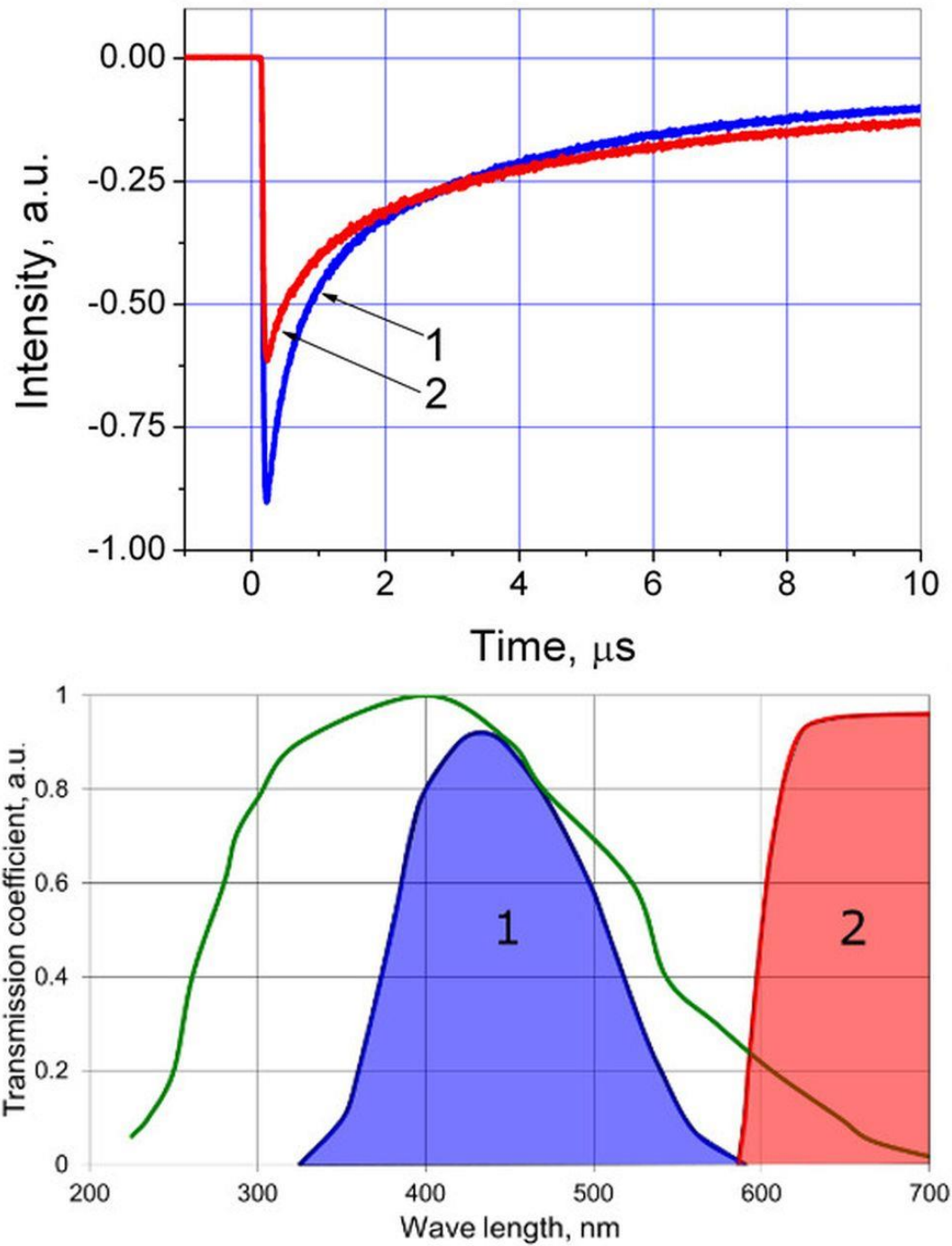

*Figure 6. a – the thermal emission accompanied the tungsten coagulation registered through the "blue" (1) and "red" (2) filters, correspondingly; b - transmission spectra of "blue" (1) and "red" (2) filters; the normalized spectral sensitivity of the PMT cathode presented in Figure as well.*

Both signals are of close amplitudes, but since the energetic width of the "red" filter is four times less than that of "blue" one and the efficiency of the PMT in the "red" region is four times lower than that in the "blue", it can be concluded that over 90% of the energy is concentrated in the red wing of emission spectrum. This is consistent with the approximation of tungsten protoclusters by a blackbody with a temperature not exceeding 4000 K. The comparison of the curves 1 and 2 temporal behavior clearly shows that at small times the blue part of the spectrum is relatively more intense than later. This is an obvious evidence of the emitter cooling down during the emission, however, as it follows from the above estimates, this effect is rather small.

But if so, the characteristic times of radiation signals presented in Figure 4 reflect mainly the kinetics of metal clusters coagulation in superfluid helium. To estimate the rate of such coagulation it makes sense first of all to take advantage of a simplified representation of the coagulation as a simple bimolecular process of two identical clusters coalescence described by the kinetic equation:

$$\frac{dn}{dt} = -kn^2 \qquad (3)$$

giving the hyperbolic temporal dependence of the reagent concentration

$$n = \frac{n_0}{1+kn_0 t} + C \qquad (4)$$

The coefficient C is introduced to remove the background connected mainly with scattering light of the room illumination. As seen in Figure 7, for the all metals decrease in the intensity of emission fits very well to the hyperbolic dependence characteristic for the recombination.

For molybdenum and especially for platinum some deviations exist for small times. We have expected such effects as manifestation of the fact that very small clusters arising on the first stage of coagulation have non-metallic binding. In this case they do not contain free electrons and their thermal emission should be much weaker than it follows from Planck formula for the blackbody.

The coefficients corresponding to the extrapolations shown in Figure 7 have the following meanings: n0 is the initial concentration of the atoms, and $k_0$ – has the sense of rate constant of the bimolecular reaction of two clusters sticking together. Since the reagents captured to the vortex can move only along the vortex core the reaction should be considered as one-dimensional and the rate constant for different metals should represent the ratio of the velocities of their motion inside the vortex. It follows from Figure 7 that the ratios of "rate constants"

$k_{Mo}$ (42):$k_W$ (74): $k_{Pt}$ (78) = 4.9: 3.7: 2.3 (atomic weights are in brackets) do not contradict this interpretation - the lighter is the atom, the faster is its motion along the core of the vortex.

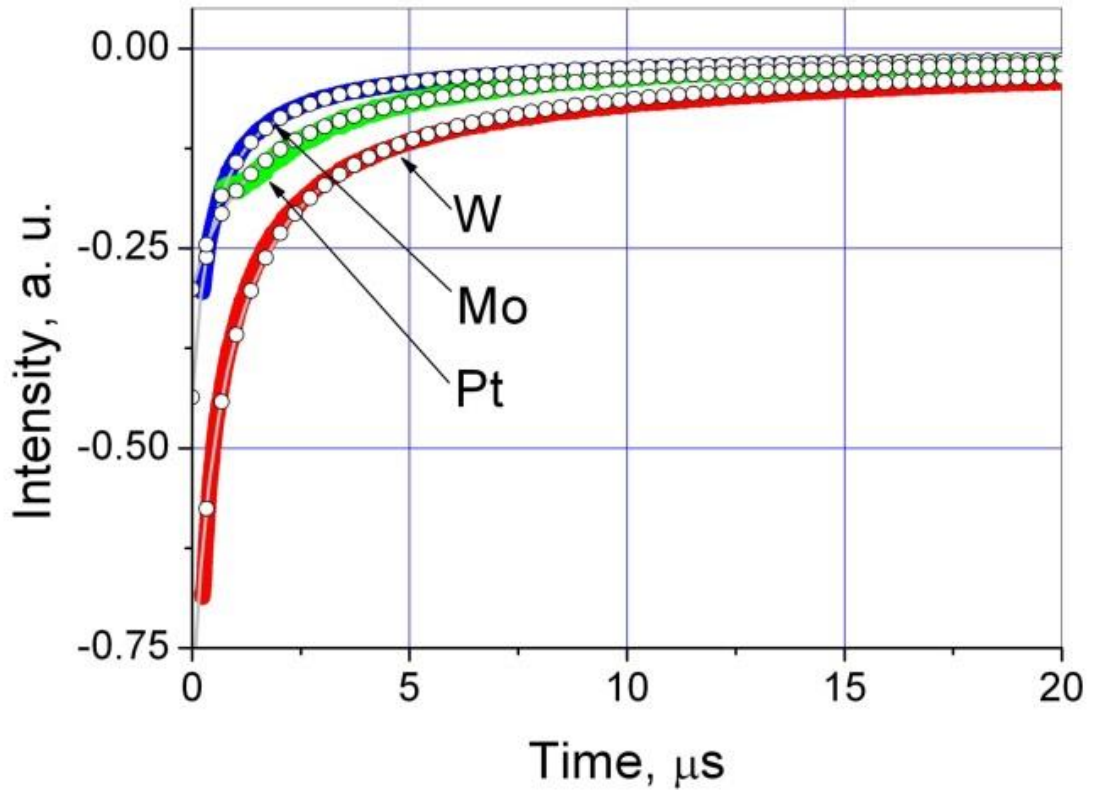

*Figure 7. The approximation of the thermal emissions temporary dependencies by the hyperbolic law for tungsten, platinum and molybdenum. Solid curves are an experimental results, open circles represent hyperbolic dependencies.*

Because of the hyperbolic dependence on time there is no concept of characteristic time for coagulation process but, as shown in Fig. 7, under conditions of our experiments the process lasts about 2 microseconds.

**6 Conclusions**

The entire dataset related to the production of nanowires by condensation of different metals inside superfluid helium have demonstrated, that the structure of nanowires (qualitatively) and their diameters (quantitatively) confirm the scenario of metal coagulation in quantized vortices of He II proposed in [16]. According to this scenario, during the initial "hot" stage the small metal clusters, even if they were previously cooled to low temperature, fuse to larger clusters and simultaneously heat above its melting temperature. Only starting from the clusters with diameter exceeded its limiting value, determined by thermo-chemical properties of the metal, they stick together into nanowires.

According to this scenario, the nanoparticles heated above their melting temperature should exist during the initial stages of metal condensation in He II. Provided they grow up to nanometer size such clusters acquire metallic binding and thus contain a lot of free electrons (especially when they are heated up to high temperatures). For that reason they should intensively emit photons in IR, and even in visible spectrum. Such radiation has been experimentally detected in the bulk He II during the condensation of tungsten, molybdenum and platinum. The spectral region of this emission shows that it belongs to the clusters heated up to at least the metal melting point. At the same time, the fusible indium, whose laser plasma emits in the visible not less efficient than other metals under study, does not demonstrate any visible radiation during coagulation; this confirms the thermal nature of the emission.

The surprising fact of refractory metals melting in He II is the consequence of the specific character of heat transfer in superfluid helium. The He II really possesses enormously high thermal conductivity but only up to very modest values of heat flow. Above this limit the heat flow becomes rather low due to the development of turbulence. Superfluid helium then converts to the normal fluid, and then it evaporates to form a heat insulating envelope filled with low-pressure helium gas.

Kinetics of the thermal radiation decay which was registered in this study and interpreted as the kinetics of metal condensation in superfluid helium is the first indication of abnormally high rate of condensation of impurities via their one-dimensional motion towards each other in quantized vortices. It turns out that with metal concentrations created inside of He II in our experiments, the entire process of condensation takes place very close to the surface of the gas plasma bubble in the laser focus, in the presence of significant turbulence.

As shown in this study, the expansion of the metal inside liquid helium is surprisingly fast, with the velocity close to Landau velocity (despite the fact that at temperature of our experiments, T = 1.7 K, the fraction of normal, viscous component is still rather high). But even this large rate is insufficient to compete with process of coagulation.

The important conclusion of this work is as follows.

The existence of huge local overheating is the result of the unique properties of superfluid helium, and it should take place not only for metals but for other materials as well. The similar effects should be observable not only in the bulk liquid helium, but also in cold helium droplets

[2,7,8,15,16]. If liquid helium droplet contained guest atoms, molecules and, especially, clusters the overheating during their chemical or photochemical processes could result in the appearance of gas cavity in the droplet. Indeed the high temperature of metal cluster in HeII exists, as follows from our results, for time of about $10^{-6}$ sec, which is few orders of magnitude longer than the time of sound wave propagation in liquid helium from the center of droplet to its surface. Because of the zero pressure around of droplet the gas cavity in the droplet should greatly expand for that time. Also the existence of the radiative (without helium atoms evaporation) channel of cluster cooling leads to underestimating of the final size of droplet.

And finally, the local overheating should occur not only in the case of chemical reactions between the particles embedded into He II, but also during coalescence of the chemically inert particles. Indeed van der Waals forces are weaker than chemical ones only by about of 30 times. It means that one can encounter there the local overheating of 30-100 K, which is significant at cryogenic temperatures.

So the existence of huge local overheating ceases any promises to produce in special conditions of superfluid helium the piece consisted of some exotic chemical compounds. However, it simultaneously gives rise to the hope to synthesize in He II unique nano-materials, the exotic properties and high possible cost of which justify the application of expensive and small-scale method of their preparation.

**Acknowledgements**

The authors are grateful to E.V. Dyatlova, A.S. Gordienko and M.E. Stepanov for participating in the experiments.

This work was financially supported by Russian Science Foundation (grant № 14-13-00574).